\documentclass[twocolumn,pra,aps,showpacs]{revtex4}

\usepackage{psfrag,graphicx}

\usepackage{dcolumn}
\usepackage{amsmath,amssymb}
\usepackage{bm}
\usepackage[dvips]{color}

\definecolor{mygrey}{gray}{0.35}
\definecolor{mygreen}{rgb}{0.85,1,0.9}
\definecolor{myzard}{cmyk}{0,0,0.05,0}
\definecolor{mywhite}{rgb}{1,1,1}
\definecolor{myred}{rgb}{1,0,0}

\def\vec#1{{\bf{#1}}}

\begin{document}

\title{Systematic Analysis of Majorization in Quantum Algorithms}

\author{Rom\'an Or\'us$^{\dag}$, Jos\'e I. Latorre$^{\dag}$ 
and Miguel A. Mart\'{\i}n-Delgado$^{\ddag}$}
\affiliation{$^{\dag}$Dept. d'Estructura i Constituents de la Mat\`eria,
Univ. Barcelona, 08028. Barcelona, Spain. \\
$^{\ddag}$Departamento de F\'{\i}sica Te\'orica I, Universidad Complutense,
28040. Madrid, Spain.}

\begin{abstract}

Motivated  by  the  need   to  uncover  some  underlying  mathematical
structure of  optimal quantum computation, we carry  out a systematic
analysis of a wide variety  of quantum algorithms from the majorization
theory point of view.  We conclude that step-by-step majorization 
is found in the known instances of fast and efficient algorithms, namely in the 
quantum Fourier transform, in Grover's algorithm, in the hidden affine 
function problem, in searching by quantum adiabatic evolution and in
deterministic quantum
walks in  continuous time solving  a classically hard problem.  On the
other hand, the optimal quantum algorithm for parity determination,
which does not provide any computational speed-up, does not 
show
step-by-step majorization. Lack of both speed-up and step-by-step majorization 
is also a feature of the adiabatic quantum algorithm solving the $2$-SAT
``ring of agrees'' problem. Furthermore, the 
quantum algorithm for the hidden affine function
problem does not make use of any entanglement while it does
obey majorization.  All the above results give support 
to a step-by-step Majorization Principle necessary for 
optimal quantum computation. 

\end{abstract}


\maketitle

\section{Introduction}

One of the main open problems in quantum computation theory is
finding some mathematical structure underlying optimal quantum
algorithms. There is a rather short list of ideas
behind the design of fast algorithms.   
Grover's quantum searching algorithm \cite{grover} exploits
calls to an oracle by enhancing a particular state
{\sl via} a rotation in
the relevant Hilbert space associated to the problem. Shor's
quantum factoring algorithm  \cite{shor} exploits the 
periodicity of an initial quantum state using only a minimum of
Hadamard gates at the core of the quantum Fourier transform.
Based on more general quantum mechanical principles, the idea
of using adiabatic evolution to carry quantum computation \cite{adiabatic}  
has proven suitable for performing Grover's algorithm
and has been numerically studied as a candidate for
attacking NP-complete problems. More recently, 
deterministic random walks in continuous time have
been proven to solve a classically hard problem in
polynomial time \cite{qwalk}. Many other quantum algorithms 
can be mapped to the above families and, therefore, bring 
no further insight.

Some attempts to uncover an underlying principle,
common to all known optimal algorithms,  have already been explored
though not definite and satisfactory answer
has been found. One relevant instance  is the
role of entanglement  in quantum algorithms \cite{ent1,ent2,ent3,ent4,rmp,guifre}.
Although entanglement is a natural resource to
 be exploited in quantum algorithm
design, there are known examples of fast algorithms
where the quantum register remains in a product state all
along the computation. Our work will concentrate on quite
a different proposal. The basic idea is that efficiency is
related to a strong majorization principle. We shall
investigate the way the probability distribution
associated to the computational basis
evolves along optimal quantum computations and find that
it obeys a very constrained behavior we shall analyze in detail.

Let us recall that majorization theory arises as the natural
 framework to study the
measure of disorder for classical probability distributions
\cite{muirhead,hardy,marshall,maj}. The
notion of ordering emerging from majorization is far more severe
than the one quantified by the standard Shannon entropy. If one
probability distribution majorizes another, a set of inequalities
must hold to constrain the former probabilities with respect to
the latter. These inequalities lead to entropy ordering, but the
converse is not necessarily true. In quantum mechanics, it has
been proven that majorization is at the heart of the solution of a
large number of quantum information problems. Majorization plays a
fundamental role in topics like ensemble realization, conversion
of quantum states via local operations and classical
communication, and characterization of positive operator valued
measurements \cite{vid}.

In the context of quantum algorithms, 
 a majorization principle has been formulated,
proven for Grover's algorithm and verified for 
Shor's algorithm in Ref. \cite{lat}. Furthermore, a 
complete proof of majorization in quantum phase-estimation algorithms
was presented in  \cite{orus}. The underlying
idea behind these analysis is that the
probability distribution associated to the quantum state in the
computational basis is step-by-step majorized until it is maximally
ordered. Then a measurement will provide the sought solution with
high probability. It has also been proven that the way such
a detailed majorization emerges in both algorithmic families is
intrinsically different \cite{orus}.

In this paper we analyze the consistency and universality
of a possible Majorization Principle. More concretely,
we have studied
several different quantum algorithms based on distinct properties. 
First, we have considered the problem of finding a hidden affine function
by means of calls to an oracle. This problem is relevant because
it is a known fast quantum algorithm that uses no
entanglement at all. Second, we have taken a non-efficient
instance, namely the parity determination problem. This is
a case where the final solution must match a global majorization,
yet it does not obey step by step majorization neither 
presents any quantum speed-up. The third case
considered here is the large class of quantum adiabatic 
evolution algorithms. Efficiency and optimality has been proven to depend
on the interpolating time path taken along the evolution. 
It is a remarkable fact that optimality appears when
step-by-step majorization is present. Finally, deterministic
quantum random walks provide exponential speed-up
over classical oracle based random walks. Again, a  
manifest strong majorization arrow is detected. 

The conclusion of the accumulated research is that all
the analyzed instances of quantum algorithms support
a step-by-step Majorization Principle. That is, optimal
quantum computation is systematically verified to 
correspond to a step by step detailed reordering of the whole
probability distribution in the computational basis. Some
of the instances show the extra feature that the initial
state can be prepared in different ways. Then, an initial 
step-by-step reverse majorization period precedes the subsequent 
step-by-step majorization, closing an invertible cycle.
The study of quantum computation by adiabatic evolution
shows the possibility of slower algorithms that maintain
majorization. This implies that step-by-step majorization
may be a necessary but is definitely not sufficient condition for
efficiency.

We have structured the paper as follows: in Sec. II we briefly
review some concepts about majorization theory and how it is
related to quantum algorithms. We develop an analysis of a quantum
algorithm for solving a hidden affine function problem in Sec.
III. In Sec. IV we study an optimal quantum algorithm solving the
parity problem. We move to an investigation of adiabatic quantum
computation in Sec. V, analyzing the effect of the
evolution path in adiabatic searching algorithms in 
Sec. V.A and Sec. V.B, as well as the
effect of the speed in the time variation of the Hamiltonian in
Sec. V.C. A further example of adiabatic evolution solving the $2$-SAT ``ring of agrees'' problem is provided in Sec. V.D. In Sec. VI we examine a recently proposed quantum
algorithm based on a continuous time quantum walk on a graph
solving a classically hard problem. Finally, in Sec. VII we
state a Majorization Principle based on all the previous results
and collect our conclusions.

\section{Majorization theory and its relation to quantum algorithms}

Our approach to the mathematical study of quantum processes will be
through majorization's eye. We review in this section the contact
point between majorization theory and quantum algorithms, as
stated previously in \cite{lat} and \cite{orus}. In particular, we
also present here the concept of ``natural majorization'', first
stated in \cite{orus}, which will eventually be used in this work.

\subsection{Majorization theory}

Let us consider two  vectors  $\vec{x}$, $\vec{y}\in \mathbb{R}^d$
such that $\sum_{i = 1}^d x_i = \sum_{i=1}^d y_i = 1$, whose
components represent two different probability distributions. We
say that distribution $\vec{y}$ majorizes distribution $\vec{x}$,
written as $\vec{x}\prec \vec{y}$, if and only if there exist a
set of permutation matrices $P_j$ and probabilities $p_j$ such
that

\begin{equation}
\vec{x} = \sum_j p_j P_j \vec{y} \ .
\label{defone}
\end{equation}
Because the probability distribution $\vec{x}$ can be obtained
from $\vec{y}$ by means of a probabilistic sum, the definition
given in equation (\ref{defone}) provides the intuitive notion
that the $\vec{x}$ distribution is more disordered than $\vec{y}$.
An alternative and usually more practical definition of
majorization can be stated in terms of a set of inequalities to be
held between the two distributions. Consider the components of the
two vectors sorted in decreasing order, written as  $(z_1, \ldots
z_d) \equiv \vec{z}^\downarrow$. We say that $\vec{y}^\downarrow$
majorizes $\vec{x}^\downarrow$ if and only if the following
relations are satisfied:

\begin{equation}
\sum_{i=1}^k x_i \leq \sum_{i=1}^k y_i \qquad k = 1 \ldots d \ .
\label{deftwo}
\end{equation}
In this paper we call probability sums similar to the ones
appearing in the previous expression as ``cumulants''. There is
still a third way of defining majorization involving the use of
doubly stochastic matrices. A real $d \times d$ matrix $D =
(D_{ij})$ is said to be doubly stochastic if it has non-negative
entries, and each row and column of $D$ sums to $1$. We say that
$\vec{y}$ majorizes $\vec{x}$ if and only if there is a doubly
stochastic matrix $D$ such that

\begin{equation}
\vec{x} = D \vec{y} \ .
\label{defthree}
\end{equation}
The equivalence among the three given definitions can be proven
\cite{maj}. Complementarily, we say that the probability
distribution $\vec{x}$ reversely majorizes distribution $\vec{y}$ if and
only if $\vec{y}$ majorizes $\vec{x}$. A powerful relation involving 
majorization and the common Shannon entropy $H(\vec{x}) \equiv -\sum_{i=1}^d 
x_i \log{x_i}$ of probability distribution $\vec{x}$ is that if $\vec{x} 
\prec \vec{y}$ then $H(\vec{y}) \geq H(\vec{x})$.

\subsection{Link between majorization theory and quantum
algorithms}

The way we relate majorization theory to quantum algorithms can be stated
as follows: let $|\psi^{(m)} \rangle$ be the pure state
representing the register in a quantum computer at an operating
stage labeled by $m = 1 \ldots M$, where $M$ is the total number
of steps in the algorithm, and let $N$ be the dimension of the
Hilbert space. If we denote as $\{|i\rangle\}_{i=1}^N$ the basis
in which the final measurement is performed in the algorithm, we
can naturally associate a set of sorted probabilities $p_i$, $i =
1 \ldots N$, to this quantum state in the following way: decompose
the register state in the measurement basis such that

\begin{equation}
|\psi^{(m)} \rangle = \sum_{i = 1}^{N}a_i^{(m)}|i \rangle \ .
\label{decomp}
\end{equation}
The probability distribution associated to this state is

\begin{equation}
\vec p^{(m)}=\{p_i^{(m)}\}\qquad
 p_{i}^{(m)} \equiv |a_{i}^{(m)}|^2 = |\langle i | \psi^{(m)} \rangle
|^2  
\label{probabs}
\end{equation}
where $i = 1 \ldots N$, which corresponds to the probabilities of all the possible outcomes if the computation were stopped at stage $m$ and a measurement were performed. A quantum algorithm will be said to majorize this probability
distribution between steps $m$ and $m+1$ if and only if \cite{lat,orus}

\begin{equation}
\vec p^{(m)} \prec \vec p^{(m+1)} \ .
\label{majflow}
\end{equation}
Similarly, a quantum algorithm will be said to reversely majorize this
probability distribution between steps $m$ and $m+1$ if and only
if

\begin{equation}
\vec p^{(m+1)}  \prec \vec p^{(m)} \ .
\label{minflow}
\end{equation}

If the situation given in equation (\ref{majflow}) is
step-by-step verified, there is a net flow of probability
directed to the values of highest weight, in such a way that the
probability distribution will be steeper as  time flows.
In physical terms, this can be stated as a very
particular constructive interference behavior, namely, a
constructive interference that has to step-by-step satisfy the
constraints given in equation (\ref{deftwo}). The quantum
algorithm builds up the solution at each time step by means of
this very precise reordering of probability distribution.

It is important to note that majorization is checked on
a particular basis. Step-by-step majorization is
a basis dependent concept. Nevertheless there is 
a preferred basis, which is the basis defined by the
physical implementation of the quantum computer
or computational basis. The principle we analyze
is rooted in the physical possibility to 
arbitrarily  stop the computation at any time and 
perform a measurement. The claim pursued along the 
paper is that the probability distribution associated
to this physically meaningful action obeys majorization.

\subsection{Natural majorization in quantum algorithms}

Let us now define the concept of natural majorization for quantum
algorithms as it was originally presented in \cite{orus}.
Working with probability amplitudes in the basis
$\{|i\rangle\}_{i=1}^N$, the action of a particular
unitary gate at step $m$ makes the amplitudes evolve to
step $m+1$ in the following way:

\begin{equation}
a_i^{(m+1)} = \sum_{j=1}^N U_{ij} a_j^{(m)} \ ,
\label{amp}
\end{equation}
where $U_{ij}$ are the matrix elements in the chosen basis of the
unitary evolution operator (namely, the propagator from step $m$
to step $m + 1$). Inverting the evolution, we can write

\begin{equation}
a_i^{(m)} = \sum_{j=1}^N C_{ij} a_j^{(m+1)} \ ,
\label{amp2}
\end{equation}
where $C_{ij}$ are the matrix elements of the inverse unitary evolution (which is unitary as well). Taking the square modulus we find

\begin{equation}
|a_i^{(m)}|^2 = \sum_{j=1}^N |C_{ij}|^2 |a_j^{(m+1)}|^2 + {\rm interference \ terms} \ .
\label{mquad}
\end{equation}
Should the interference terms disappear, majorization would be
verified in a ``natural'' way between steps $m$ and $m+1$  because
the initial probability distribution could be obtained from the
final one only by the action of a doubly stochastic matrix with
entries $|C_{ij}|^2$. This is the so-called ``natural
majorization'': majorization which naturally emerges from the
unitary evolution due to the lack of interference terms when
making the square modulus of the probability amplitudes.
Similarly, we can define the concept of ``natural reverse majorization'',
which follows in a similar way: there will be ``natural
reverse majorization'' between steps $m$ and $m+1$ if and only if there is
``natural majorization'' between time steps $m+1$ and $m$.

\subsection{Majorization in Grover's and Shor's quantum algorithms}

In order to motivate the forthcoming study we briefly sketch here 
the main results found concerning the analysis 
of majorization in the two well-known quantum algorithms
of Grover \cite{grover} and Shor \cite{shor}. These results were 
already presented in Ref. \cite{lat} and \cite{orus}, so we address
the reader interested in precise details to these references.

On the one hand, Grover's quantum searching algorithm was found
in \cite{lat} to follow a step-by-step majorization. 
More concretely, each time the Grover's
operator is applied the probability distribution obtained 
from the computational basis obeys the set of constraints given in 
equation (\ref{deftwo}) until the searched state is found. Furthermore,
because of the possibility of understanding Grover's quantum evolution as
a rotation in a two-dimensional Hilbert space (see for example
\cite{nielsen}) it is seen that the quantum algorithm follows 
a step-by-step reverse majorization when evolving far away from 
the marked state, until the initial superposition of all possible 
computational states is obtained again. The quantum algorithm behaves
 such that majorization is present when approaching to 
the solution, while reverse majorization appears when escaping from it. A cycle of
majorization and reverse majorization emerges in the process as we consider
long enough time evolutions, 
due to the rotational nature of Grover's operator.

On the other hand, Shor's quantum algorithm was analyzed inside of 
the broad family of quantum phase-estimation algorithms. In \cite{lat} 
it was observed that a step-by-step majorization seemed to appear under
the action of the last quantum Fourier transform when considered
in the usual Coppersmith decomposition \cite{copper}. One step further 
was taken in \cite{orus}, were the complete mathematical proof of this
property was provided. The result relies on the fact that those quantum 
states that can be mixed by a Hadamard operator coming from the decomposition
of the quantum Fourier transform only differ by a phase all
along the computation. Such a property entails as well the appearance
 of natural majorization, in the way presented above.   
Natural majorization was proven relevant for the case of 
Shor's quantum Fourier transform. This particular algorithm
manages step-by-step majorization in a most efficient
way. No interference terms spoil the majorization introduced
by the natural diagonal terms in the unitary evolution. It
is still unclear the role that natural majorization plays in
distinguishing different level of efficiency in quantum
algorithms.

These two results suggest a possible relation between majorization
and quantum algorithms. This is the point to be exploited in detail with
further examples in the next sections.

\section{Analysis of a quantum algorithm for solving a hidden affine function problem}

The problem of finding hidden affine functions was initially
proposed by Bernstein and Vazirani \cite{BZ} as a generalization
of Deutsch's problem \cite{deutsch}. Further studies have
investigated into this class of problems, providing a range of
fast quantum algorithms for solving different generalizations  
 \cite{cleve,mosca}. The case we present in
this work is one of the multiple variations stated in
Ref. \cite{mosca}, but the main results we obtain can
be verified as well for the whole set of quantum
algorithms solving similar problems.

\subsection{Setting of the problem}

Let us state the following problem (see \cite{mosca}):

\bigskip

\emph{Given an integer N function $f:x \rightarrow mx+b$, where
$x, m, b \in \mathbb{Z}_N$, find $m$.}

\bigskip

The classical analysis reveals that no information about $m$ can
be obtained with only one evaluation of the function $f$.
Conversely, given the unitary operator $U_f$ acting in a
reversible way in the Hilbert space $\mathbb{H}_N \otimes
\mathbb{H}_N$ such that

\begin{equation}
U_f |x\rangle |y\rangle = |x\rangle |y + f(x)\rangle \ ,
\label{f}
\end{equation}
(where the sum is to be interpreted as modulus $N$), there is a
quantum algorithm solving this problem with only one query to
$U_f$.

\subsection{Quantum algorithm}

Let us take $N=2^n$, being $n$ the number
of qubits. The quantum algorithm optimally solving the problem
previously presented reads as follows:

\begin{itemize}

\item{ Prepare two registers of $n$ qubits in the state $|0 \ldots 0 \rangle |\psi_1 \rangle \in \mathbb{H}_N \otimes \mathbb{H}_N$, where $|\psi_1 \rangle = QFT(N)^{-1}|1\rangle$, and $QFT(N)^{-1}$ denotes the inverse quantum Fourier transform in a Hilbert space of dimension $N$.}

\item{ Apply $QFT(N)$ over the first register.}

\item{ Apply $U_f$ over the whole quantum state.}

\item{ Apply $QFT(N)^{-1}$ over the first register.}

\item{ Measure the first register and output the measured value.}

\end{itemize}

\begin{figure}[h]
\centering
\includegraphics[angle=0, scale=0.4]{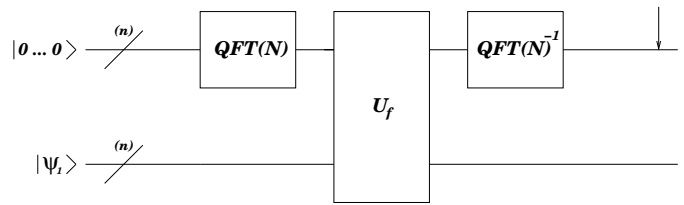}
\caption{Quantum circuit solving the hidden affine function problem. Each
quantum wire is assumed to be composed of $n$ qubits. The arrow at
the end indicates a measurement.} 
\label{circ}
\end{figure}

The different steps concerning this process are summarized in Fig. \ref{circ}.

\bigskip

\subsection{Analysis of the quantum algorithm}

We now show how the proposed quantum algorithm leads
to the solution of the problem. Our analysis raises  two
observations concerning the way both entanglement and majorization
behave in the computational process.

In the first step of the algorithm, the quantum state is 
separable, noting that the quantum Fourier transform (and
its inverse) applied 
on 
a well defined state in the
computational basis leads to a perfectly separable state (see, for
example, 
\cite{cleve})
Actually, this separability holds also
step-by-step when a decomposition for the quantum Fourier
transform is considered, such as the Coppersmith's
decomposition \cite{copper}. That is, the quantum state $|0\ldots
0\rangle |\psi_1\rangle$ is unentangled.

The second step of the algorithm corresponds to a
quantum Fourier transform in the first register.
This action leads to a step-by-step reverse majorization of the probability
distribution of the possible outcomes while it does
not use neither create any entanglement. Moreover,  
natural reverse majorization is at work due to the absence of interference terms.

Next, it is easy to verify that the quantum state
\begin{equation}
|\psi_1 \rangle = \frac{1}{\sqrt{N}}\sum_{j=0}^{N-1}e^{-2 \pi \frac{j}{N}}|j\rangle
\label{secreg}
\end{equation}
is an eigenstate of the operation $|y\rangle \rightarrow |y +
f(x)\rangle $ with eigenvalue $e^{2 \pi i \frac{f(x)}{N}}$. After
the third step, the quantum state reads
\begin{equation}
\frac{1}{\sqrt{N}}\sum_{x = 0}^{N-1}e^{2 \pi i \frac{f(x)}{N}}|x\rangle |\psi_1 \rangle = \frac{e^{2 \pi i \frac{b}{N}}}{\sqrt{N}} \left( \sum_{x=0}^{N-1}e^{2 \pi i \frac{mx}{N}}|x\rangle \right) |\psi_1 \rangle \ .
\label{equation}
\end{equation}
The probability distribution of possible outcomes has not been
modified, thus not affecting majorization. Furthermore, the pure
quantum state of the first register can be written as
$QFT(N)|m\rangle$ (up to a phase factor), so this step has not
created any entanglement among the qubits of the system either.

In the fourth step of the algorithm, 
the action of the operator $QFT(N)^{-1}$ over the
first register leads to the state $e^{2 \pi i
\frac{b}{N}}|m\rangle |\psi_1\rangle$. A subsequent  measurement in the
computational basis over the first register provides the desired
solution. Recalling the results found in \cite{orus}, we see that
the inverse quantum Fourier transform naturally majorizes
step-by-step the probability distribution attached to the
different outputs. On the other hand, the separability of the
quantum state still holds step-by-step. Note that  step-by-step majorization 
is in fact dependent on the specific implementation of the quantum Fourier 
transform operation, but nevertheless it holds true for other decompositions 
of the operator appart from the usual Coppersmith's one (as we already stated in \cite{orus}). 

It is clear that the quantum algorithm is faster than any of its
possible classical counterparts, as it only needs of a
single query to the unitary operator $U_f$ to get the solution. We can 
summarize our analysis of majorization for the present
quantum algorithm as follows:

\bigskip
{\bf Result 1:}  \emph{ The fast quantum algorithm 
for finding a hidden affine function
shows a majorization cycle based on the action of a
$QFT(N)$ and a $QFT(N)^{-1}$.}
\bigskip

We understand that the algorithm is entanglement-free as long as we analyze it
between the action of the different unitary gates. From a more physical point
of view, the quantum registers may become highly entangled during the performance of
multi-qubit gates, despite it is not present between two of them. Our assertion
relies then on the study of the system at these particular steps in the
computation, which we think to be the most natural steps to consider.  
It follows that there can exist quantum
computational speed-up 
without the use of entanglement (in the way made precise before). In this case, it is seen
that no 
resource increases exponentially. Yet, a
majorization cycle is present in the process, which is indeed 
rooted in the structure of both the quantum Fourier transform 
and the quantum state.

\section{Analysis of an optimal quantum algorithm for solving the parity problem}

The problem of finding the parity of a given function $f: x \in
\mathbb{Z}_N \rightarrow \{-1, +1\}$, usually known as the parity
problem, has been shown to be as hard for a quantum computer in
the quantum oracular model of computation as it is for a classical
computer \cite{parity1,parity2}: no quantum speed-up can be
achieved in this case. We shall first present the problem and
then analyze an optimal quantum algorithm proposed
in \cite{parity1} from the point of view of
majorization.

\subsection{Setting of the problem}

Let us state the parity problem in the following way (see \cite{parity1}):

\bigskip
\emph{Given a function $f:x \in \mathbb{Z}_N \rightarrow \{-1,
+1\}$, evaluate the product of all the $f(x)$ for all the possible
$x$.}
\bigskip

It has been proved that a  quantum computer will need at least
$N/2$ queries to the quantum oracle $f$  for solving this problem
compared to the $N$ classical queries \cite{parity1}. Thus, a
quantum computer is not faster than a classical one (in the limit of a very large input, where $N$ goes to infinity)
when dealing
with this situation: no better efficiencies can be obtained using the
quantum computational paradigm in getting the desired result. The time complexity of the best possible quantum algorithm will be $O(N)$, without improvement with respect to the classical time complexity, because the quantum speed-up is only by a factor of two.

\subsection{Quantum Algorithm}

Let us  briefly outline the main lines of an optimal quantum
algorithm solving the parity problem in exactly $N/2$ queries to
the oracular function $f$, which was initially presented in
\cite{parity1}. We first introduce a series of definitions and
notations: the function $f(x)$ will be evaluated through a quantum
oracle acting in the following way
\begin{eqnarray}
& U_f |x, +1 \rangle = |x , f(x)\rangle \nonumber \\
& \nonumber \\
& U_f |x, -1 \rangle = |x, -f(x)\rangle \ , \label{oracle}
\end{eqnarray}
where the second register is a qubit taking the values $\pm 1$.
Let us define  
also
the quantum state $|x, a\rangle =
\frac{1}{\sqrt{2}}\left(|x, +1\rangle - |x, -1\rangle \right)$,
which is seen to be a proper state of operator $U_f$ with
eigenvalue $f(x)$. With this definitions, the quantum algorithm
reads as follows:

\begin{itemize}
\item{Prepare the initial quantum state $|\psi_0\rangle = \frac{1}{\sqrt{N}}\sum_{x = 1}^{N}|x, a\rangle $.}
\item{Apply the following operations over the initial quantum state:

\begin{equation}
V_{N/2} U_f V_{N/2-1} U_f \cdots V_1 U_f |\psi_0 \rangle \equiv |\psi_f \rangle \ ,
\label{operations}
\end{equation}
where $U_f$ is defined as before, and the rest of operators are
defined as $ V_{N/2} = 1 $, $ V_i = V \ \forall \ i \neq N/2 $,
with

\begin{equation}
\begin{split}
V |x, a\rangle &= |x + 1, a \rangle \ \ x = 1, \ldots, \frac{N}{2} -1  \\
V |\textstyle{\frac{N}{2}}, a \rangle &= |1, a\rangle  \\
V |x, a \rangle &= |x+1, a\rangle \ \ x = \frac{N}{2}+1, \ldots, N-1  \\
V |N, a\rangle &= |\frac{N}{2} + 1, a\rangle \ .
\end{split}
\label{unitary}
\end{equation}
}

\item{Measure the observable $|\psi_0 \rangle \langle \psi_0 |$ over $|\psi_f \rangle$ .}

Note that the final measurement is to be made on a specific 
basis.

\end{itemize}

\subsection{Analysis of the quantum algorithm}

We now make an study of how this algorithm leads to the desired
solution of the proposed problem. The analysis will show the 
way majorization behaves in
this optimal but non-efficient quantum process.

We focus on how the operations in the second step affect the
quantum state leading to the solution, and thus affect the
probability distribution of possible outcomes for the final
measurement. If we apply $U_f$ to the initial state, we get

\begin{equation}
U_f |\psi_0 \rangle = \frac{1}{\sqrt{N}}\sum_{x = 1}^{N} f(x) |x,a\rangle \ .
\label{eqs}
\end{equation}
After the application of operator $V_1$ the quantum state evolves to

\begin{eqnarray}
V_1 U_f |\psi_0 \rangle && = \frac{1}{\sqrt{N}}
\sum_{x=1}^{N/2}f(x-1)|x,a\rangle \nonumber \\
&& + \frac{1}{\sqrt{N}}\sum_{x = N/2
+1}^{N}f(x-1)|x, a\rangle \ , \label{step1}
\end{eqnarray}
as can be directly checked
(care must be taken with the possible values of
$x$ in both sums).
If we now apply again $U_f$ we get:

\begin{eqnarray}
U_f V_1 U_f |\psi_0 \rangle && = \frac{1}{\sqrt{N}}\sum_{x =
1}^{N/2}f(x)f(x-1) |x, a \rangle \nonumber \\
&& + \frac{1}{\sqrt{N}}\sum_{x =N/2+1}^{N}f(x)f(x-1)|x, a \rangle \ , \label{evolvution}
\end{eqnarray}
so we begin to recognize the pattern the algorithm follows. At the
end of the computation the final state is

\begin{eqnarray}
|\psi_f\rangle && = \frac{1}{\sqrt{N}} f(1)\cdots
f(N/2)\sum_{x=1}^{N/2}|x, a\rangle \nonumber \\
&& + \frac{1}{\sqrt{N}}f(N/2 + 1)\cdots
f(N)\sum_{x=N/2 + 1}^{N}|x, a\rangle \ , \label{final}
\end{eqnarray}
and it is easily verified that this state is equal to $|\psi_0
\rangle$ if the parity of the function is equal to $+1$, and
orthogonal to $|\psi_0 \rangle$ (say $|\psi_0 ^{\bot}\rangle $) if
the parity is $-1$. A suitable measurement of the observable $|\psi_0
\rangle \langle \psi_0 |$ can then distinguish between the two
values.

Let us now analyze the way this algorithm behaves with respect to
majorization. As stated in Sec. II.B, majorization must always
be checked from the probability distribution of obtaining the
different outcomes of the final measurement. In other words, the
probability distribution subject of analysis must always be the
one obtained from the final measurement basis. Such a basis turns
usually to be the computational one, but it is not necessarily so.
Here, we are dealing with one of these exceptional cases in which the final
measurement basis differs from the computational one.
Consequently, majorization must be studied in this unusual but
natural basis for the quantum algorithm.

 The only two vectors we know of this basis are $|\psi_0\rangle$
and $|\psi_0^{\bot}\rangle$. We could extend them to a whole basis
but it is not necessary for our purposes, as we can analyze the
probability of being in each of these two states all along the
computation. Should majorization be present step-by-step in the
process, the probability of being in one of these two states would
smoothly decrease in favor of the other one, which would
parallely smoothly increase. In a naive way, this is what a
\emph{majorization arrow} means: as the process evolves the
probability of being in the right state becomes bigger and bigger.

We can observe that this does not happen in the algorithm for the
whole computational process, except for the last application of
the oracle $U_f$ (compulsory if one wishes to distinguish between
the two states). It is easily seen, because in all the steps of
the computation the quantum state is an arbitrary superposition of
computational states of amplitudes $+1/\sqrt{N}$ and
$-1/\sqrt{N}$, without any apparent structure, so the
probabilities of being in $|\psi_0 \rangle$ or
$|\psi_0^{\bot}\rangle$ evolve erratically. The full 
structure only appears
when the last oracle $U_f$ is applied, thus providing the
necessary majorization to be able to distinguish the two states
with a measurement, but the important point is that this is not a
majorization arrow, because there is no step-by-step majorization.
This is stated in our second result:

\bigskip
{\bf Result 2:} \emph{ No step-by-step majorization is
present along the non-efficient parity determination problem.}
\bigskip

Thus, no majorization cycle similar to the one found in the
preceding section could ever appear.
Interestingly enough,
this is a
problem in which quantum computers do not provide a better efficiency
than classical ones.

\section{Analysis of the adiabatic searching algorithms and of an adiabatic algorithm solving a $2$-SAT problem}

We now turn to the  analysis of the
quantum adiabatic searching algorithm, observing the effects of a
change of path between the initial and the problem hamiltonian
under the majorization's point of view \cite{lat}. We see that those paths
leading to optimality in the algorithm lead as well to
step-by-step majorization, while the converse is not necessarily
true. A different example of adiabatic evolution is analyzed in the last point 
of this section, namely, an adiabatic algorithm solving the $2$-SAT ``ring of agrees'' problem.

The adiabatic model of quantum computation was initially introduced
in \cite{adiabatic}, and can be briefly summarized as follows. We
consider a physical system controlled by a 
a time dependent hamiltonian of the form

\begin{equation}
H(s(t)) = (1-s(t)) H_0 + s(t) H_p \ ,
\label{ham}
\end{equation}
where $H_0$ and $H_p$ are the initial and problem hamiltonian
respectively, and $s(t)$ is a time dependent function satisfying
the boundary conditions $s(0)=0$ and $s(T) = 1$ for a given $T$.
The desired solution to a given problem is 
encoded 
in the ground
state of $H_p$. The gap between the ground and the first excited
state of the instantaneous hamiltonian at time $t$ will be called
$g(t)$. Let us define $g_{min}$ as the global minimum of $g(t)$
for $t$ in the interval $[0, T]$. If at time $T$ the ground state
is given by the state $|E_0; T\rangle$, the adiabatic theorem
states that if we prepare the system in its ground state at $t=0$
(which is assumed to be easy to prepare) and let it evolve under
this hamiltonian, then

\begin{equation}
|\langle E_0; T|\psi(T)\rangle|^2 \geq 1 - \epsilon^2
\label{probab}
\end{equation}
provided that

\begin{equation}
\frac{{\rm max} |\frac{dH_{1,0}}{dt}|}{g^2_{min}} \leq \epsilon
\label{cond}
\end{equation}
where $H_{1,0}$ is the hamiltonian matrix element between the
ground and first excited state, 
$\epsilon \ll 1$,
and the 
maximization is taken over the whole time interval $[0,T]$ \cite{adiabatic,wim}.
Because the problem hamiltonian  
encodes
the solution to the
problem in its ground state, we get the desired solution with high
probability after a time $T$.

This quantum adiabatic evolution method has been 
applied to the searching problem (\cite{das,roland,wim,jer}).
 Let the initial state be
$|\psi_0 \rangle = \frac{1}{\sqrt{N}}\sum_{x = 1}^N |x\rangle$,
being $N$ the number of entries for the searching, and let the
initial and problem hamiltonian be $H_0 = I - |\psi_0\rangle
\langle \psi_0|$ and $H_p = I - |m\rangle \langle m|$, being
$|m\rangle$ the searched state. This scheme leads to different
results depending on whether we apply the adiabatic condition
globally (that is, in the whole time interval $[0,T]$) or locally
(at each time $t$). In what follows we will consider these two
situations without entering into precise details of the involved
calculations.
For further information, we refer the reader to \cite{das,roland}
and references therein.

\subsection{Analysis of the fastest global adiabatic evolution}
Let us suppose that we demand the adiabatic condition given
in equation (\ref{cond}) to be satisfied globally in the whole
interval $[0,T]$. This does not involve any particular restriction
on the form of $s(t)$, so we can then choose $s(t) = t/T$, leading
to a linear evolution of the hamiltonian. Under these
circumstances it can be proven \cite{roland} that the global
adiabatic condition is verified provided that
\begin{equation}
T \geq \frac{N}{\epsilon} \ .
\label{timmme}
\end{equation}
Hence, the algorithm needs  $O(N)$ time to hit the solution with
appreciable probability, so the global adiabatic searching does
not lead to an increasing efficiency with respect to a classical
searching (in contrast with the square root speed-up of Grover's
quantum algorithm).

Let us call $P_+(t)$ the probability of being at the searched
state at time $t$ 
and similarly $P_-(t)$ the probability of being at any
different state from the desired one at time $t$.
Note
that, because of the symmetry of the problem, $P_-(t)$ will be
exactly the same for all those quantum states differing from the
solution state all along the computational process. In order to
analyze majorization, we recall the set of inequalities given in
equation (\ref{deftwo}) to be satisfied at each majorizing time
step. It is easy to see that the maximum probability at all times
will be $P_+(t)$, while the rest of the probabilities will remain
smaller to this quantity all along the computation and equal to
$P_-(t)$. In order to gain simplicity we have analyzed in detail 
the behaviour of the two non-trivial cumulants $P_+(t)$ and 
$P_+(t) + P_-(t)$, as the rest of them will not lead to different 
conclusions from the ones emerging from our study.  

\begin{figure}[h]
\centering
\includegraphics[angle=-90, width=0.5\textwidth]{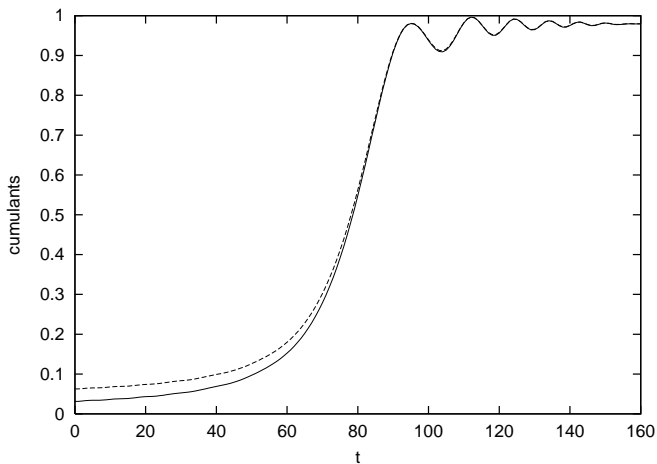}
\caption{Global adiabatic evolution with parameters
$\epsilon = 0.2$, $N = 32$ and $T = 160$.
The solid line corresponds to the time evolution of
$P_+(t)$ and the dashed line that of $P_+(t) + P_-(t)$.} 
\label{adi1}
\end{figure}

We have performed numerical simulations in the fastest allowed
case ($T = \frac{N}{\epsilon}$) and have found the time 
evolution for the two cumulants. The results for $\epsilon = 0.2$
and $N = 32$ are shown in Fig. \ref{adi1}.

From our numerical analysis we arrive at our third result:

\bigskip
{\bf Result 3:} \emph{A naive adiabatic quantum searching
process does not  produce an optimal algorithm neither verifies  
 step-by-step majorization}.
\bigskip

This property is clearly seen as the two cumulants \emph{decrease}
in time for some time steps, thus not verifying
majorization. 

\subsection{Analysis of the local adiabatic evolution}

The preceding global adiabatic method can be improved if we apply
the adiabatic condition given in equation (\ref{cond}) locally.
That is, let us divide the interval $[0, T]$ into many
subintervals and let us apply (\ref{cond}) to each of the
subintervals individually. Taking the limit of the size of the
subintervals going to zero, we find that the adiabatic restriction
has to be fulfilled locally at each time $t$:

\begin{equation}
\frac{|\frac{dH_{1,0}}{dt}|}{g^2(t)} \leq \epsilon \qquad \forall t \ .
\label{cond2}
\end{equation}
This is a less demanding condition than (\ref{cond}) (if \ref{cond} is
satisfied, so is \ref{cond2}, but the inverse is not necessarily true), and
means that the adiabatic evolution must be satisfied at each
infinitesimal time interval. It can be shown (see, for example,
\cite{roland}) that proceeding in this way the function $s(t)$
must have a precise form which is given by the relation

\begin{figure}[h]
\centering
\includegraphics[angle=-90, width=0.5\textwidth]{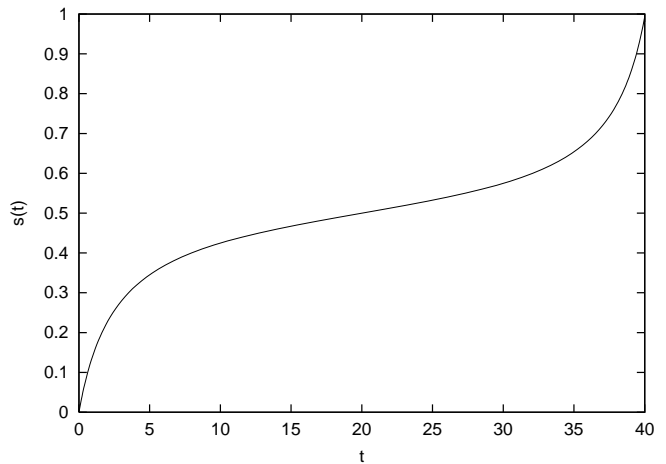}
\caption{Local adiabatic evolution. The driving hamiltonian, with $\epsilon = 0.2$ and $N = 32$.}
\label{esefig}
\end{figure}

\small
\begin{equation}
t = \frac{1}{2 \epsilon} \frac{N}{\sqrt{N-1}} \left(
\arctan({\sqrt{N-1}(2s-1)}) + \arctan({\sqrt{N-1}})\right) \ .
\label{te}
\end{equation}
\normalsize
We can observe this dynamic evolution from Fig. \ref{esefig}, in
the case of $\epsilon = 0.2$ and $N = 32$.
The local adiabatic
process implies that the smaller the energy gap between the
ground and first excited states, the slower the evolution of
the hamiltonian (and viceversa).

With this information it can be proven as well \cite{roland}
that the evolution
time for the algorithm to succeed with appreciable probability is, in
the limit where $N \gg 1$,
\begin{equation}
T = \frac{\pi}{2 \epsilon}\sqrt{N} \ .
\label{time}
\end{equation}
Hence, in the case of local adiabatic evolution the computational
process takes $O(\sqrt{N})$ time, just as in Grover's quantum
searching algorithm, obtaining an square root speed-up with
respect to a classical searching.

\begin{figure}[h]
\centering
\includegraphics[angle=-90, width=0.5\textwidth]{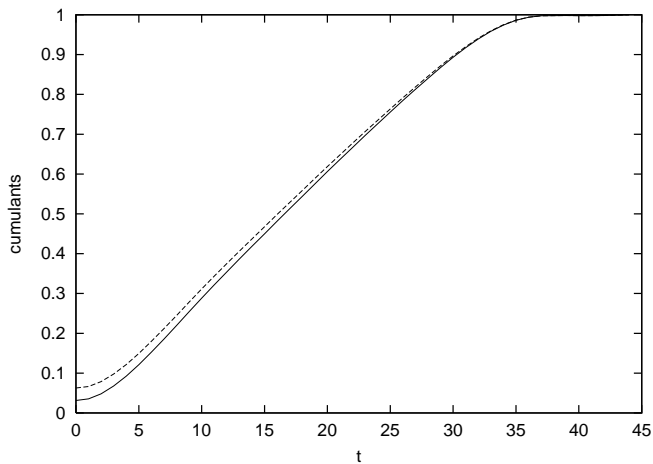}
\caption{Local adiabatic evolution with parameters
$\epsilon = 0.2$, $N = 32$ and $T = 44$. 
The solid line corresponds to the time evolution of
$P_+(t)$ and the dashed line that of $P_+(t) +
P_-(t)$.} 
\label{adi2}
\end{figure}

Defining again $P_+(t)$ and $P_-(t)$
in the same way as in Sec. V.A,
we can restrict ourselves to the analysis of the two
non-trivial cumulants $P_+(t)$ and $P_+(t) + P_-(t)$ in order to
observe the evolution of majorization in the quantum process. We
have numerically solved the equations for $\epsilon = 0.2$ and $N
= 32$, and have found the evolution of the two quantities, which
is given in Fig. \ref{adi2}.

From the numerical analysis the following result emerges:

\bigskip
{\bf Result 4:} \emph{A local adiabatic 
searching algorithm is optimal and verifies  step-by-step majorization}.
\bigskip

This result can be straightforwardly verified since the set of
inequalities of (\ref{deftwo}) are  satisfied step-by-step,
according to the plot in Fig. \ref{adi2}.
Due to this
behavior, \emph{the whole computational process might lead to a
majorization cycle}, such as the one observed in Sec. III, as
long as the previous preparation of the initial quantum state of
the computation at time $t=0$ leads to a step-by-step
reverse majorization. This turns to be always possible, for example, by
applying a set of Hadamard gates to the $\log{N}$-qubit quantum
state $|0\ldots 0\rangle$ (assume that $N$ is a power of $2$, for
simplicity), which would efficiently prepare a superposition of
all the possible quantum states together linked to a reverse majorization
arrow. Furthermore, this quantum process leads to an increasing
efficiency with respect to a classical searching, exactly in the
same fashion as Grover's original quantum searching algorithm \cite{grover}.

\subsection{Analysis of slower global adiabatic evolutions}

Let us now address the situation of global adiabatic
evolutions which are not necessarily tight in time, that is,
extremely slow time variations for the hamiltonian, much 
slower than the minimum necessary for the adiabatic theorem to
hold. In the case we are dealing with, that implies the
consideration  of the case in which $T > \frac{N}{\epsilon}$, i.e., 
the adiabatic inequality (\ref{timmme}) is
not tight. This case is not very relevant from a computational
point of view because the hitting time is not the minimum
possible, but we think it is worthwhile to be studied also from
the point of view of majorization theory in order to have a
more complete picture of how majorization really works in this
kind of quantum algorithms.

\begin{figure}[h]
\includegraphics[angle=-90,
width=0.5\textwidth]{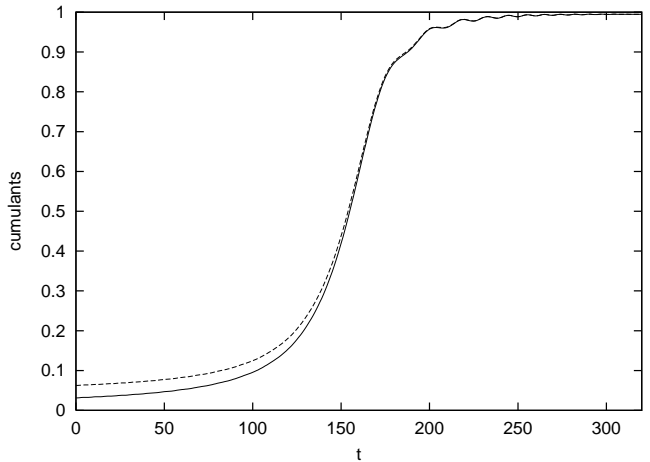}
\caption{ Global adiabatic evolution with  $\epsilon = 0.2$, $N = 32$, and
 $T = 320$. The solid line
corresponds to the time evolution of $P_+(t)$ and the dashed line
that of $P_+(t) + P_-(t)$.} 
\label{adi34}
\end{figure}

\begin{figure}[h]
\includegraphics[angle=-90, width=0.5\textwidth]{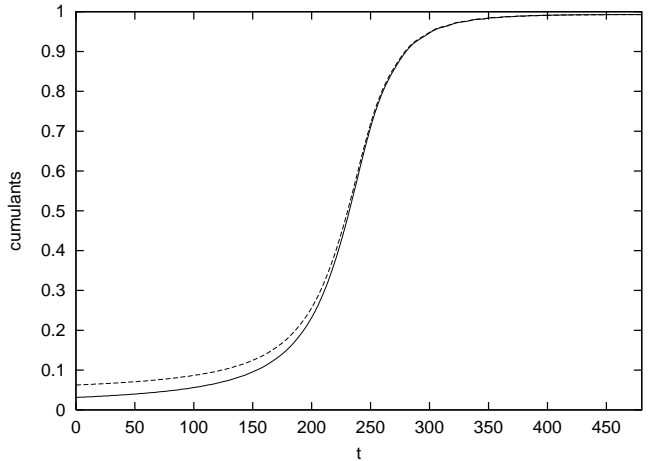}
\caption{ Global adiabatic evolution with $\epsilon = 0.2$, $N = 32$, and $T =
  480$. The solid 
line corresponds to the time evolution of $P_+(t)$ and the dashed line that of $P_+(t) + P_-(t)$.}
\label{adi35}
\end{figure}

We have performed again numerical analysis for the time evolution
of the two non-trivial cumulants, for $\epsilon = 0.2$, $N = 32$,
and $T = 320, 480$ (in both cases bigger than $\frac{N}{\epsilon}
= 160$). The results are plotted in Fig. \ref{adi34} and Fig. \ref{adi35}.

From these two plots, we observe with Fig. \ref{adi1}, 
that a step-by-step majorization appears as long as the
evolution of the hamiltonian becomes slower and slower.
Physically, this means that the probability of ``jumping'' to the
first excited state diminishes as long as the evolution is
performed with a very small velocity, thus satisfying better the
assumptions of the adiabatic theorem.
This leads to processes in
which there is no quantum speed-up but there is indeed 
a majorization arrow. However, these processes are not optimal in
time, as we can always find faster quantum algorithms for solving
the problem. Consequently, we arrive at the following conclusion:

\bigskip
{\bf Result 5:} \emph{Non-optimal quantum algorithms may
present step-by-step majorization. In particular, step-by-step majorization may
appear in global adiabatic searching processes for a slow enough evolution
rate.}
\bigskip

It follows that step-by-step majorization cannot be a sufficient 
condition for quantum speed-up.

We can get some further intuition of the set of results presented in this
section.
Adiabatic quantum searching algorithms can be understood (in the limit
of large $N$) as a rotation from the initial state to the marked state
as long as the adiabaticity (either global or local) of the evolution
is conveniently satisfied (see \cite{jer} for details). The difference
between the global and local conditions turns out to be the evolution
rate of the rotation angle: local adiabatic evolution imposes a
rotation at constant rate (as in the original Grover's algorithm)
whereas global adiabatic evolution does not. Because of this
rotational picture, step-by-step majorization is verified as long as
the quantum state remains in the instantaneous ground state all along
the computation. We can now understand our results in a finer
way. Global adiabatic evolution is not a
strong enough condition for adiabaticity, thus we only see step-by-step
majorization when the evolution is \emph{really} slow, in which case
the quantum state adiabatically rotates towards the solution because
it remains very close to the ground state of the instantaneous
hamiltonian. Local adiabatic evolution is a
stronger condition for adiabaticity the quantum state remains always
very close to the instantaneous ground state, thus performing the
rotation towards the solution which in turn involves step-by-step
majorization.

\subsection{A further example: a $2$-SAT quantum adiabatic
 algorithm solving the ``ring of agrees'' problem}

Let us consider now a different example of adiabatic quantum
computation, namely, an adiabatic quantum algorithm solving the
$2$-SAT ``ring of agrees'' problem, as stated in \cite{adiabatic}. As
long as $2$-SAT can be efficiently solved by a classical algorithm in
a time $O({\rm poly}(n))$ (being $n$ the number of bits) \cite{2sat},
quantum computation can do no better than classical computation in
this case. The problem hamiltonian $H_p$ is now a sum of hamiltonians
involving each of the different clauses of the $2$-SAT problem,
whereas the initial hamiltonian $H_0$ is such that its ground state is
again an equal superposition of all the possible states of the
computational basis. The ``ring of agrees'' problem over $n$ bits is
defined such that clause $j$ acts on bits $j$ and $j+1$ where $j$ runs
from $1$ to $n$ and bit $n+1$ is identified with bit $1$. Each clause
is an ``agree'' clause, which means that $00$ and $11$ are the
satisfying assignments. The eigenvectors of the hamiltonian associated
with clause $j$ are the computational states, in such a way that those
which ``agree'' in qubits $j$ and $j+1$ have zero energy (ground
states) whereas those which ``disagree'' have energy one. Because the
problem hamiltonian is a sum of the $n$ hamiltonians for the $n$
``agree'' clauses, its ground states are
$|0\rangle|0\rangle\cdots|0\rangle$,
$|1\rangle|1\rangle\cdots|1\rangle$, or any linear combination of
them.

\begin{figure}[h]
\includegraphics[angle=-90, width=0.5\textwidth]{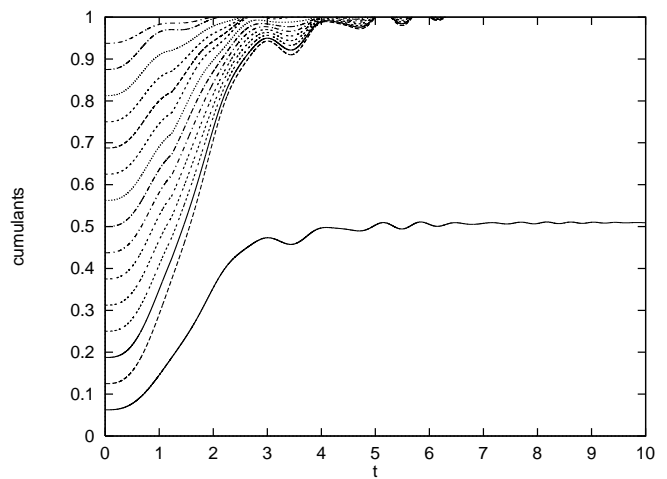}
\caption{Evolution of the 15 cumulants in the ``ring of agrees'' for a
Hilbert space of dimension 16}
\label{sat}
\end{figure}

We have made an analysis of the adiabatic quantum algorithm in which
the interpolation between the initial and problem hamiltonian is
linear, $s(t) = t/T$, for the case of $4$ qubits (Hibert space of
dimension $16$), and choosing $T=10$. The evolution of the $15$
cumulants is plotted in Fig. \ref{sat}.

Note the similarity between Fig. \ref{sat} and Fig. \ref{adi1} in
Sec. V.A, namely, no step-by-step majorization is present in the
evolution because of the oscilatory behaviour of the cumulants. Both
plots represent quantum algorithms which do not improve classical
computation and which probabilities share the same behaviour under the
point of view of majorization. Our observation is then the following:

\bigskip
{\bf Result 6:} \emph{A quantum adiabatic algorithm solving the
$2$-SAT ``ring of agrees'' problem does not improve classical
computation, neither verifies step-by-step majorization.}
\bigskip

This result reinforces the ones already found with respect to
adiabatic searchig algorithms.

\section{Analysis of a quantum walk in continuous 
time with exponential algorithmic speed-up}

The extension of classical random walks to the quantum world has
been widely studied, yielding  two different models of
quantum random walks, namely, those which operate in discrete time
by means of using a ``coin operator'' \cite{coin1,coin2,coin3}
 and those  based on a
hamiltonian evolution in continuous time \cite{cont1,cont2,qwalk}.
Regarding the discrete time model of quantum random walk two
interesting algorithmic results have been found so far, namely, an
exponentially faster hitting time in the hypercube with respect to the
classical random walk \cite{kempe1} and a quantum searching algorithm
achieving the Grover's quadratic speed-up \cite{kempe2}. The first of
these examples does not provide any algorithmic speed-up, as there exists
 a classical algorithm that solves the hitting problem in the
hypercube exponentially faster than the naive classical random walk,
that is, in a time $O({\rm poly}(\log{N}))$ where $N$ is the number of
nodes of the graph (see \cite{qwalk}). On the other hand, the second of these examples
shows algorithmic advantage with respect to any possible classical
computation. The analysis of  the quantum random walk searching
algorithm
shows that  the quantum evolution can be understood as an (approximate)
rotation of the quantum state in a two-dimensional Hilbert space which
is exact in the limit of a very large database (see \cite{kempe2} for
details), resembling the original proposal of Grover's searching
algorithm which can be decomposed exactly in a two-dimensional Hilbert
space. This rotational structure of the evolutin implies step-by-step
majorization when approaching the marked state, exactly in the same
fashion than the usual Grover's searching algorithm (already analyzed
in \cite{lat}).

In this section we restrict ourselves to the continuous time 
model of quantum walk and analyze a recently proposed quantum
algorithm based on a quantum walk on continuous time solving a
classically hard problem \cite{qwalk}. Here we restrict ourselves
to briefly sketch the main points and ideas of both the problem
setting and its quantum efficient solution, 
since
the whole
development of the algorithm is not the purpose of the present
paper. We address the interested reader to \cite{qwalk}.

\subsection{Setting of the problem}

\begin{figure}[]
\centering
\includegraphics[angle=0, width=0.403\textwidth]{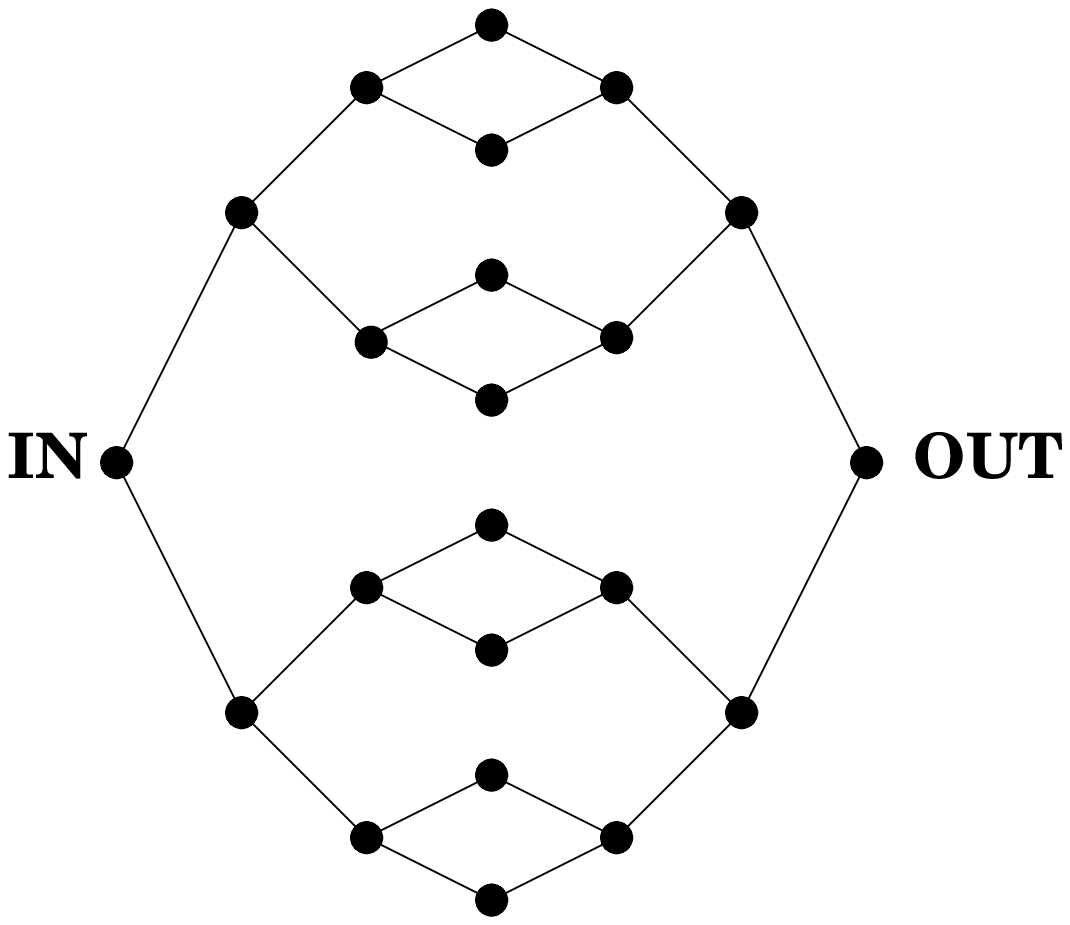}
\caption{A possible graph constructed from two binary trees with $n = 3$.}
\label{ge}
\end{figure}

\begin{figure}[]
\centering
\includegraphics[angle=0, width=0.5\textwidth]{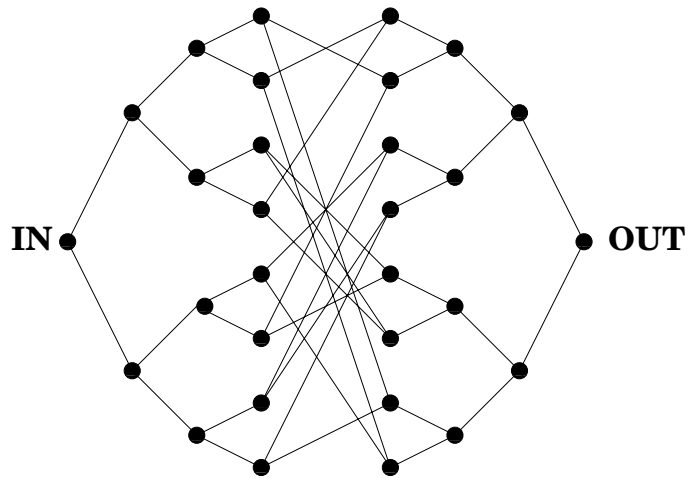}
\caption{An alternative graph constructed from two binary trees
with $n = 3$ (connection between the leaves is made through a
random cycle).} \label{geprime}
\end{figure}

The problem we wish to solve is defined through a graph built in the
following way (see \cite{qwalk}): suppose we are given two 
balanced binary trees of height $n$ with the $2^n$ leaves 
of the left tree identified with
the $2^n$ leaves of the right tree in a simple way, as shown in
Fig. \ref{ge}. A way of modifying such a graph is by connecting
the leaves by a random cycle that alternates between the leaves of
the two trees, instead of identifying them directly. An example of
such a graph is provided in Fig. \ref{geprime}.

Suppose that the edges of such a graph are assigned a consistent
coloring (that is, not two edges incident in the same vertex have
the same color), and that the vertices are each one given a
different name (with a $2n$-bit string, so there are more possible
names than the ones assigned). We now define a black box that
takes two inputs, a name $a$ given as a $2n$-bit string and a
color $c$, and acts in the following way: if the input name $a$
corresponds to a vertex that is incident with an edge of color
$c$, then the output corresponds to the name of the vertex joined
by that edge; if $a$ is not the name of a vertex or $a$ is the
name of a vertex but there is no incident edge of color $c$, the
output is the special $2n$-bit  string $11\ldots 1$, which is not
the name of any vertex.

Now, the problem we wish to solve reads as follows:

\bigskip

\emph{Given a black box for a graph such as the one previously
described, and given the name of the {\rm IN} vertex, find the
name of the {\rm OUT} vertex.}

\bigskip

In \cite{qwalk} it was proven that no classical algorithm can
transverse a graph such as the one in Fig. \ref{geprime} in
polynomial time, given such a black box. Furthermore, 
an explicit contruction of
a quantum algorithm based on a continuous time quantum walk on the
graph that succeeds in finding the solution for this oracular
problem in polynomial time was given.

\subsection{Quantum algorithm}

The quantum algorithm of \cite{qwalk} can be briefly summarized as
follows: consider the $(2n+2)$-dimensional subspace spanned by the
states

\begin{equation}
|{\rm col} \ j \rangle = \frac{1}{\sqrt{N_j}}\sum_{a \in  {\rm column} \  j}|a\rangle \ ,
\label{column}
\end{equation}
where $N_j = 2^j$ if $0 \leq j \leq n$ and $N_j = 2^{2n+1-j}$ if
$n+1 \leq j \leq 2n+1$. We will call this subspace the ``column
subspace'', and each state of the basis is an equally weighted sum
of the states corresponding to the vertices lying on each column
of the graph. We now define a hamiltonian acting on this subspace
in the following way:

\small
\begin{equation}
\langle {\rm col} \ j |H|{\rm col} \ (j+1)\rangle = \Big\{
\begin{array}{ll}
 1 & 0 \leq j \leq n-1 \ , \ n+1 \leq j \leq 2n   \\
 \sqrt{2} & j = n   \\
\end{array}
\label{chamiltonian}
\end{equation}
\normalsize
with hermiticity of $H$ giving the other nonzero matrix elements.
The action of this hamiltonian in the graph is nothing but
promoting transitions between adjoint vertices, so a quantum walk
on the graph (on the whole Hilbert space) generated by this
hamiltonian is equivalent to a quantum walk on the line (on the
column subspace). Consequently, from now on we only focus our
attention in the quantum walk in the line generated by the
hamiltonian of (\ref{chamiltonian}). Moreover, it  can be
proven that given the structure of the graph in the form of a
black box such as the one already described, our hamiltonian can
be efficiently simulated \cite{qwalk}.

The quantum walk works as follows: at first the ``wave packet''
will be precisely localized at the IN vertex (the initial state
will be $|{\rm col} \ 0\rangle$). Due to unitary time evolution, it
will initially spread out through the different vertices at the
left hand side of the graph (those belonging to the left binary
tree), but after a short time (once half the graph has been
transversed) it will begin to spread through the vertices on the
right hand side, interfering constructively in the OUT vertex as
the time goes on. Physically, this is nothing but a wave
propagation. Should we wait more time, the wave packet would come
back to the entrance, and the process would be similarly repeated
again. Actually, due to the ``defect'' of the hamiltonian in the
middle vertices, it can be shown that the transmission through the
central columns is not of $100$ percent (thus providing
interferences in long enough time scales), but high enough for the
OUT node to be achieved with a very high probability. In
\cite{qwalk} the authors prove that the succeeding time is
polynomial in $n$.

\subsection{Analysis of the quantum algorithm}

\begin{figure}[h]
\centering
\includegraphics[angle=-90, width=0.5\textwidth]{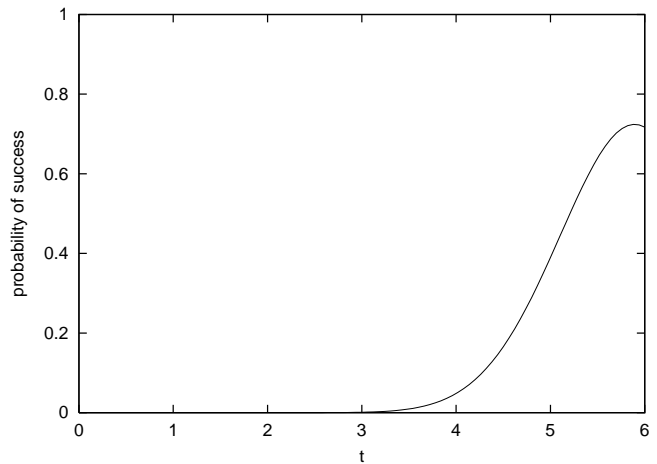}
\caption{Quantum walk algorithm. Probability of finding the exit, for $n=4$.}
\label{prob}
\end{figure}

We have numerically simulated this quantum walk for the particular
case of $n =4$, and have plotted the time evolution of the
probability of success in Fig. \ref{prob}. We observe that the
numerical result fits with the prediction that the time
the algorithm takes in achieving the OUT node is polynomial.

\begin{figure}[h]
\centering
\includegraphics[angle=-90, width=0.5\textwidth]{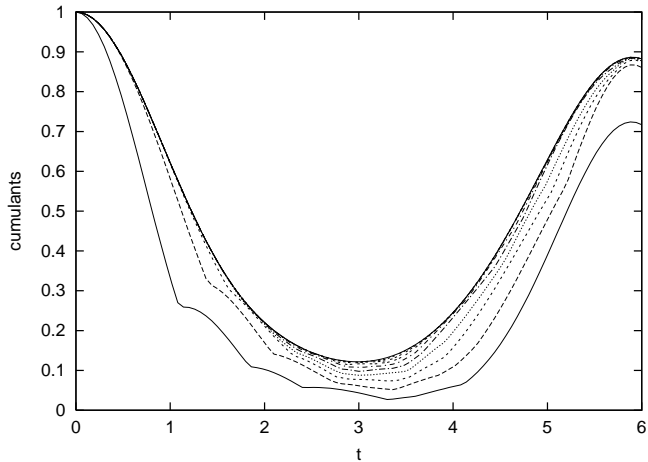}
\caption{Quantum walk algorithm. Time evolution of the ten cumulants when one node per column is considered, for $n=4$.}
\label{cycle}
\end{figure}

It is easy to observe that, in order to analyze majorization, for
the case of $n=4$ there are only $10$ non-trivial probabilities to be
studied. This is so due to the fact that all the states of the
whole Hilbert space belonging to the same column always share the
same probability amplitude. The relevant quantities to be studied
are then the probabilities of being at each column state
normalized to the number of nodes belonging to that column, that
is, the probability of being in one node of each column. There are
then $2n+2$ different probabilities to be considered at each time
step. Given only this $10$ quantities, we are able to calculate the whole set of $62$ cumulants corresponding to all the sums of sorted probabilities, according with equation (\ref{deftwo}). In order to make the figures as clear as possible we only plot $10$ of these quantities in Fig. \ref{cycle}, corresponding to the cumulants arising from the sorted probabilities when only one node per column is considered. The rest of the cumulants can be shown to have a similar behavior to that of the ones appearing in Fig. \ref{cycle}.  
 
\begin{figure}[h]
\centering
\includegraphics[angle=-90, width=0.5\textwidth]{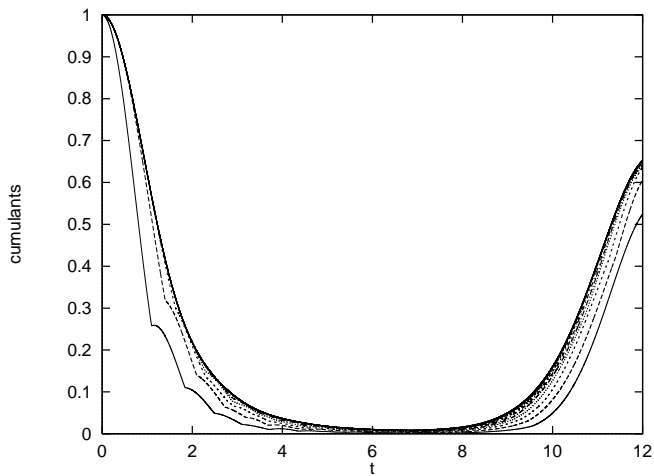}
\caption{Quantum walk algorithm. Time evolution of the $22$ cumulants when one node per column is considered, for $n=10$. Note that most of the quantities apparently collapse when making the plot because of the small difference in the probabilities given the big size of the graph.}
\label{cycle2}
\end{figure}

We have also numerically simulated the algorithm in the case of a bigger graph, namely, in the case $n=10$. In this case there are $2n+2 = 22$ different probabilities to be considered at each time step. Proceeding in the same way than in the case $n=4$ (that is, not plotting all the cumulants, but the only the sorted sum of these 22 probabilities), we obtain a similar behaviour as in the case for $n=4$, as is shown in Fig. \ref{cycle2}.

Therefore, we arrive at the following conclusion:

\bigskip
{\bf Result 7:} \emph{The continuous time quantum walk transversing
 a classically hard graph follows a step-by-step majorization
cycle all along the computation until it reaches the {\rm OUT}
node}.
\bigskip

It is worth remarking as well that the time the algorithm spends
reversely majorizing the probability distribution is about half of the time
of the whole computation. The physical reason for this behavior
is clear, as this is the time the ``wave packet'' spends spreading
over the binary tree on the left hand side, thus leading to a
destructive interference part. It is note worthy that such a
destructive interference indeed strictly follows a step-by-step
reverse majorization of probabilities (satisfying the inequalities given
in equation (\ref{deftwo}) for the case of reverse majorization).
Furthermore, we see combining Fig. \ref{prob} and Fig. \ref{cycle}
that the growing of the probability of success is linked
to a step-by-step majorization. Physically, this is the part in
which the algorithm constructively interferes into the OUT node
once the wave packet is approximately in the right hand side binary
tree. We see that this constructive interference follows a
majorization arrow, thus 
verifying  step-by-step
the inequalities
given in equation (\ref{deftwo}). Actually, the observed
majorization cycle reminds us the one already found in the quantum
algorithm of Sec. III, but in this case we have numerically checked
that the present cycle does not seem to follow the rules of 
natural majorization. Complementarily, we have also observed 
that the probability amplitudes follow the interesting rule that
those belonging to even columns are real and those belonging to 
odd columns are imaginary. 

\begin{figure}[h]
\centering
\includegraphics[angle=-90, width=0.5\textwidth]{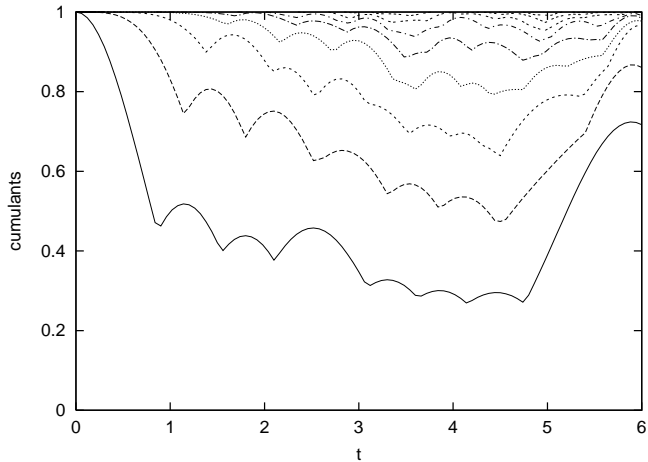}
\caption{Quantum walk algorithm. Time evolution of the nine cumulants when the column measurement is considered, for $n=4$.}
\label{nocycle}
\end{figure}

\begin{figure}[h]
\centering
\includegraphics[angle=-90, width=0.5\textwidth]{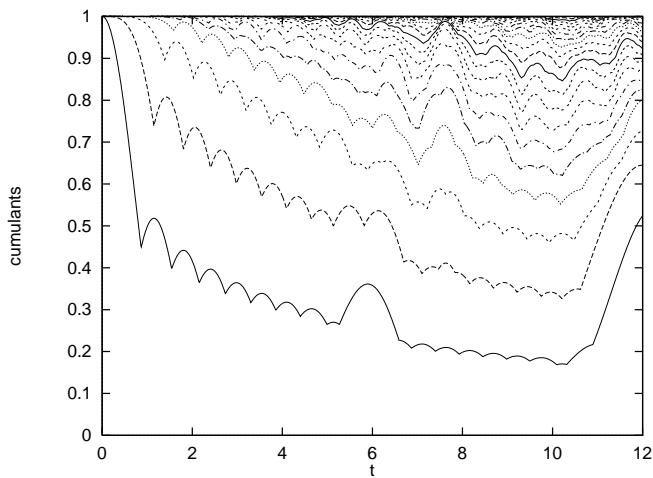}
\caption{Quantum walk algorithm. Time evolution of the $21$ cumulants when the column measurement is considered, for $n=10$.}
\label{nocycle2}
\end{figure}

The deterministic search by quantum random walk heavily exploits
the column structure of the problem. The register works
on a superposition of columns, that is of states belonging
to the same column with equal weight. It is then natural
to ask whether a step-by-step majorization cycle operates
also at the level of columns. The idea behind this
analysis corresponds to accept that the final measurement will
filter each columns as a whole. The result of the measurement 
would correspond to determining a particular column. The subtle point
here is to find to what extend the success of finding the
OUT state is related to the column structure of the algorithm.   
We have numerically considered the column amplitudes
 for $n=4$  and $n=10$ with a total of
 $9$ and $21$ cumulants to be calculated respectively from the sorted 
probabilities at each time step of being \emph{at each column} 
of the graph. In Fig. \ref{nocycle} and Fig. \ref{nocycle2} we plot the 
result, which shows that  \emph{there does not exist a majorization 
cycle when the final measurement is made on columns}. 

The conclusion is that deterministic quantum walks cleverly 
exploit the column subspace structure of the problem to achieve
step-by-step majorization on the individual states.

\section{Conclusion: a Majorization Principle}

We can now collect all the results found in the analysis
of majorization in the quantum algorithms we have
studied so far and synthesize
an emerging principle underlying all of them.
There are a total of nine empirical observations
about step-by-step majorization:

\begin{itemize}

\item{Presence of step-by-step majorization in Grover's quantum searching algorithm \cite{lat}.}
\item{Presence of a natural step-by-step majorization cycle in quantum phase estimation algorithms \cite{orus}.}
\item{Presence of natural step-by-step majorization cycle  in the quantum algorithm for finding hidden affine functions (Sec. III.C).}
\item{Absence of step-by-step majorization in an optimal, yet non-efficient,  quantum algorithm for solving the parity problem (Sec. IV.C).}
\item{Absence of step-by-step majorization in naive global quantum adiabatic searching algorithms (Sec. V.A).}
\item{Presence of step-by-step majorization in appropriate local quantum adiabatic searching algorithms (Sec. V.B).}
\item{Emergence of step-by-step majorization  for a slow enough
 evolution rate
 in quantum adiabatic searching algorithms (Sec. V.C).}
\item{Absence os step-by-step majorization in an adiabatic quantum algorithm solving the $2$-SAT ``ring of agrees'' problem (Sec. V.D).}
\item{Presence of step-by-step majorization cycle in a deterministic 
quantum walk on a graph solving a classically hard problem (Sec. VI.C).}

\end{itemize}

Note that our results concerning the analysis of adiabatic quantum
algorithms can have an alternative valid interpretation: according to
Fig. \ref{adi1} and Fig. \ref{sat}, we see that the part of the processes
which does not obey step-by-step majorization only occurs when the probability of
succes has almost achieved its highest value. We are then led to the
consideration that absence of majorization only appears once the algorithms have
already constructed the right solution, having already done their job. A
redefinition of the algorithm by stopping the
process once the probability of the winner is maximum would lead us to affirm 
that step-by-step majorization is naturally present in quantum algorithms by
adiabatic evolution according to the evidence presented here. This
new interpretation does not alter our final result. 

Adiabatic algorithmic processes do lead as well to a reverse majorization
of the probability distribution in order to efficiently prepare
the initial quantum state of the computation. This
can be efficiently performed by a set of Hadamard gates (which
produce step-by-step natural reverse majorization). Nevertheless, this remark does not only
hold for the adiabatic paradigm. The usual formulation of Grover's
algorithm in terms of quantum gates needs as well of a preparation
of the initial quantum state which can be carried out exactly in
the same way. Similarly, all algorithms accomodate to a
reverse majorization-majorization cycle. We could argue that the initial
step-by-step reverse majorization procedure at the beginning of the quantum
algorithms is somehow trivial, as it only involves the application of
(for instance) a set of hadamard gates (with the exception of the
quantum walk algorithm, in which the initial step-by-step reverse majorization
is by no means trivial in the sense we state here as it is carried by
the structure of the graph). The fact that
 the quantum evolution accomodates to 
a step-by-step reverse majorization-majorization cycle is reminiscent of the
reversibility of these quantum algorithms.

All the results
found so far suggest that a  step-by-step reverse majorization-majorization 
cycle seems to be a \emph{necessary} condition for efficiency in quantum
computational processes, although \emph{not sufficient}.
This can be promoted to a principle:

\bigskip

\textbf{Majorization Principle:}

\emph{Optimal quantum algorithms must follow a majorization cycle.}

\bigskip

This principle fits well with all the observed results given so far. Note 
that those processes which are not optimal do not necessarily follow the 
majorization cycle pattern: the case of the optimal algorithm solving the 
parity problem does not, while the extremely slow and  inefficient but 
majorizing adiabatic processes do. Step-by-step majorization may be viewed 
as a strong irreversibility condition for success probability necessary for 
optimal quantum algorithms.

All our results are also consistent with a stronger statement, namely
that both step-by-step majorization and large entanglement
complement each other and 
are needed for exponential speed-up. Entanglement brings
the genuine quantum mechanical tool which has to be used
in an optimal way, that is verifying step-by-step majorization.

\textbf{Acknowledgments:} We acknowledge financial support from the
projects MCYT FPA2001-3598, GC2001SGR-00065, IST-1999-11053, PB98-0685
and  BFM2000-1320-C02-01. One of us (R. O.) wishes to thank A. Childs,
E. Deotto, E. Farhi, J. Goldstone, S. Gutmann, A. Landahl and R. Sharma for
insightful comments and stimulating discussions.

{}


\begin{thebibliography}{}

%
\bibitem{grover}
L. K. Grover, Phys. Rev. Lett. \textbf{78}, 325 (1997)

%
\bibitem{shor}
P. W. Shor, Proc. 35th IEEE, Los Alamitos CA, 352 (1994);
quant-ph/9508027.

%
\bibitem{adiabatic}
E. Farhi, J. Goldstone, S. Gutmann, M. Sipser; quant-ph/0001106.

%
\bibitem{qwalk}
A. M. Childs, R. Cleve, E. Deotto, E. Farhi, S. Gutmann, D.
Spielman; quant-ph/0209131.

%
\bibitem{ent1}
J. Ahn, T. C. Weinacht, P. H. Bucksbaum; Science 287(5452),
463-465, 21 January 2000.

%
\bibitem{ent2}
P. Knight; Science 287(5452), 441-442, 21 January 2000.

%
\bibitem{ent3}
S. Lloyd; quant-ph/9903057.

%
\bibitem{ent4}
R. Jozsa, N. Linden; quant-ph/0201143.

%
\bibitem{rmp}
A. Galindo, M.A. Martin-Delgado.
Rev. Mod. Phys. {\bf 74},  347-423 (2002); quant-ph/0112105.

%
\bibitem{guifre}
G. Vidal; quant-ph/0301063

%
\bibitem{muirhead}
R. F. Muirhead, Proc. Edinburg Math. Soc. {\bf 21}, 144, (1903).

%
\bibitem{hardy}
G. H. Hardy, J. E. Littlewood, G. P\'{o}lya, \emph{Inequalities},
Cambridge University Press, 1978.

%
\bibitem{marshall}
A. W. Marshall, I. Olkin, \emph{Inequalities: Theory of
Majorization and its Applications.} Acad. Press Inc., 1979.

%
\bibitem{maj}
R. Bathia, \emph{Matrix Analysis} Graduate Texts in Mathematics
vol. 169, Springer-Verlag, 1996.

%
\bibitem{vid}
M. A. Nielsen, G. Vidal, Quantum Information and Computation, {\bf
1}, 76, (2001).

%
\bibitem{lat}
J. I. Latorre, M. A. Mart\'{\i}n-Delgado; Phys. Rev. {\bf A}66,
022305 (2002); quant-ph/0111146.

%
\bibitem{orus}
R. Or\'{u}s, J. I. Latorre, M. A. Mart\'{\i}n-Delgado;
Quantum Information Processing, {\bf 4}, 283-302 (2003); quant-ph/0206134. 

%
\bibitem{BZ}
E. Bernstein, U. Vazirani, Quantum complexity theory; SIAM Journal
on Computing, 26(5): 1411-1473, October 1997.

%
\bibitem{deutsch}
D. Deutsch; Proc. R. Soc. Lond. A, 400:97, 1985.

%
\bibitem{cleve}
R. Cleve, A. Ekert, C. Macchiavello, M. Mosca; Proc. R. Soc.
London, Ser. A \textbf{454}, 339 (1998).

%
\bibitem{mosca}
M. Mosca, \emph{Quantum Computer Algorithms} (Ph.D. Thesis).

%
\bibitem{nielsen}
M. A. Nielsen, I. Chuang, \emph{Quantum Computation and Quantum
Information.} Cambridge University Press, 2000.

%
\bibitem{copper}
D. Coppersmith; IBM Research Report Report 19642, 1994.
quant-ph/0201067.

%
\bibitem{parity1}
E. Farhi, J. Goldstone, S. Gutmann, M. Sipser; Phys.Rev.Lett. 81
(1998) 5442-5444; quant-ph/9802045.

%
\bibitem{parity2}
R. Beals, H. Buhrman, R. Cleve, M. Mosca, R. de Wolf; Proc. of the
99th Annual Symposium on Foundations of Computer Science
(FOCS'98), 352-361, Los Alamitos, California, November 1998. IEEE;
quant-ph/9802049.

%
\bibitem{das}
S. Das, R. Kobes, G. Kunstatter; quant-ph/0204044.

%
\bibitem{roland}
J. Roland, N. J. Cerf; Phys. Rev. A {\bf 65}, 042308 (2002);
quant-ph/0107015.

%
\bibitem{wim}
W. van Dam, M. Mosca, U. Vazirani; Proceedings of the 42nd Annual Symposium of Computer Science, 279-287 (2001); quant-ph/0206003.

%
\bibitem{jer}
J. Roland, N. J. Cerf; quant-ph/0302138.

%
\bibitem{2sat}
S. A. Cook; Proc. 3rd Ann. ACM Symp. on Theory of Computing, 151-158 (Association for Computing Machinery, New York, 1971). 

%
\bibitem{coin1}
Y. Aharonov, L. Davidovich, N. Zagury; Phys. Rev. A {\bf 48}, 1687
(1993).

%
\bibitem{coin2}
D. Aharonov, A. Ambainis, J. Kempe, U. Vazirani; Proc. of the 33rd
ACM Symposium on the Theory of Computing, 50 (ACM Press, New York,
2001).

%
\bibitem{coin3}
A. Ambainis, E. Bach, A. Nayak, A. Vishwanath, J. Watrous; Proc.
33rd Symposium on the Theory of Computing, 37 (ACM Press, New
York, 2001).

%
\bibitem{cont1}
E. Farhi, S. Gutmann; Phys. Rev. A {\bf 58}, 915 (1998);
quant-ph/9706062.

%
\bibitem{cont2}
A. M. Childs, E. Farhi, S. Gutmann; Quantum Information Processing,
{\bf 1}, 35 (2002); quant-ph/0103020.

%
\bibitem{kempe1}
J. Kempe; quant-ph/0205083.

%
\bibitem{kempe2}
N. Shenvi, J. Kempe, K. Birgitta Whaley; quant-ph/0210064.


\end{thebibliography}
\end{document}